\documentclass[twocolumn,aps,prd,showpacs,10pt,superscriptaddress]{revtex4-1}

\usepackage{amssymb}
\usepackage{amsmath}
\newcommand{\vect}[1]{\boldsymbol{#1}} 
\usepackage{multirow}
\usepackage{graphicx}
\usepackage[usenames,dvipsnames,svgnames]{xcolor}
\usepackage{hyperref}
\hypersetup{
pdfnewwindow=true,      
colorlinks=true,        
linkcolor=Blue,         
citecolor=Blue,         
filecolor=Blue,         
urlcolor=Blue           
}

\begin{document}

\title{Equivalence of the equilibrium and the nonequilibrium molecular dynamics methods for thermal conductivity calculations: From bulk to nanowire silicon}
\author{Haikuan Dong}
\email{dhk@bhu.edu.cn}
\affiliation{School of Mathematics and Physics, Bohai University, Jinzhou, China}
\affiliation{School of Materials Science and Engineering, Liaoning University of Technology, Jinzhou, China}
\author{Zheyong Fan}
\email{brucenju@gmail.com}
\affiliation{School of Mathematics and Physics, Bohai University, Jinzhou, China}
\affiliation{QTF Centre of Excellence, Department of Applied Physics, Aalto University, FI-00076 Aalto, Finland}
\author{Libin Shi}
\affiliation{School of Mathematics and Physics, Bohai University, Jinzhou, China}
\author{Ari Harju}
\affiliation{QTF Centre of Excellence, Department of Applied Physics, Aalto University, FI-00076 Aalto, Finland}
\author{Tapio Ala-Nissila}
\affiliation{QTF Centre of Excellence, Department of Applied Physics, Aalto University, FI-00076 Aalto, Finland}
\affiliation{Center for Interdisciplinary Mathematical Modeling and Departments of Mathematical Sciences and Physics, Loughborough University, Loughborough, Leicestershire LE11 3TU, UK}

\date{\today}

\begin{abstract}
Molecular dynamics simulations play an important role in studying heat transport in complex materials. The lattice thermal conductivity can be computed either using the Green-Kubo formula in equilibrium MD (EMD) simulations or using Fourier's law in nonequilibrium MD (NEMD) simulations. These two methods have not been systematically compared for materials with different dimensions and inconsistencies between them have been occasionally reported in the literature. Here we give an in-depth comparison of them in terms of heat transport in three allotropes of Si: three dimensional bulk silicon, two-dimensional silicene, and quasi-one-dimensional silicon nanowire. By multiplying the correlation time in the Green-Kubo formula with an appropriate effective group velocity, we can express the running thermal conductivity in the EMD method as a function of an effective length and directly compare it with the length-dependent thermal conductivity in the NEMD method. We find that the two methods quantitatively agree with each other for all the systems studied, firmly establishing their equivalence in computing thermal conductivity.
\end{abstract}

\maketitle

\section{Introduction}

The molecular dynamics (MD) simulation method is one of the most valuable numerical tools in investigating heat transport properties, especially for complex structures where methods based on lattice dynamics are computationally formidable. The equilibrium MD (EMD) method based on the Green-Kubo formula \cite{green1954jcp,kubo1957jpsj} and the nonequilibrium MD (NEMD) method \cite{ikeshoji1994mp,jund1999prb,muller1997jcp,wirnsberger2015jcp} based on Fourier's law are the two mainstream methods for computing lattice thermal conductivity in MD simulations, although the approach-to-equilibrium method \cite{lampin2013jap,melis2014epjb,zaoui2016prb,zaoui2017prb} has also become popular recently. 

A crucial difference between the EMD and the NEMD methods concerns the finite-size effects introduced by a finite simulation cell \cite{wang2017jap}. In the EMD method, when periodic boundary conditions are applied, one usually can obtain a size-independent thermal conductivity using a relatively small simulation cell and the cell size does not correspond to a real sample size as in an experimental measurement setup. In the NEMD method, the simulation cell length (in the transport direction) is supposed to be the sample length as in real experiments. Therefore, when the cell length is smaller than the overall phonon mean free path, the heat transport is partially ballistic (transporting without scattering) and the thermal conductivity should be smaller than that in an infinitely long system. Usually, due to the relatively large phonon mean free path, it is hard to directly simulate up to the length at which the thermal conductivity becomes fully converged, and one usually resorts to extrapolation to estimate the length-convergent thermal conductivity. 

A natural question is whether or not the converged thermal conductivity as obtained in the NEMD method is consistent with (within statistical errors) that calculated using the EMD method. There have been a few works focusing on the comparison between the two methods \cite{schelling2002prb,sellan2010prb,he2012pccp,howell2012jcp}. These works have mainly studied three-dimensional (3D) bulk silicon, described either by the Stillinger-Weber (SW) \cite{stillinger1985prb}  or the Tersoff  \cite{tersoff1989prb} empirical many-body potential. In the case of the SW potential, excellent agreement between the two methods have been found by Howell \cite{howell2012jcp}. However, in the case of the Tersoff potential, Howell \cite{howell2012jcp} did not attempt to make a comparison, while He \textit{et al.} \cite{he2012pccp} found that there are noticeable discrepancies between the two methods for certain simulation parameters. Significant discrepancies between the two methods have also been reported for other good heat conductors such as GaN modeled by a Stillinger-Weber potential \cite{liang2015jap}. Comparisons between the two methods have been less attempted for low-dimensional systems and discrepancies have been occasionally reported. For single-layer graphene, the thermal conductivity predicted by some EMD simulations \cite{pereira2013prb} is significantly smaller than that predicted by other NEMD simulations \cite{park2013jap}, using the same interatomic potential. For single-layer silicene \cite{ vogt2012prl, chen2012prl}, the two-dimensional (2D) allotrope of Si, it has been reported \cite{zhang2014prb} that the two methods are inequivalent. For quasi-one-dimensional (Q1D) silicon nanowire (SiNW) \cite{li2003apl}, divergent thermal conductivity (with respect to system length) has been reported \cite{yang2010nt} based on NEMD simulations, which was not supported by recent EMD simulations \cite{zhou2017nl}. Therefore, it is important to unravel the possible reasons behind the discrepancies reported between EMD and NEMD simulations.

In this work, we make detailed comparisons between the EMD and the NEMD methods in the calculation of the thermal conductivity $\kappa$ of three Si-based materials, including 3D bulk silicon, 2D silience and Q1D SiNW, using the newly developed GPUMD (Graphics Processing Units Molecular Dynamics) package \cite{fan2017cpc,fan2017gpumd}. In the EMD method, $\kappa$ is calculated as a function of the correlation time $t$ while in the NEMD method, $\kappa$ is calculated as a function of the system length $L_x$. We find that $\kappa(t \rightarrow \infty)$ from the EMD simulations and $\kappa(L_x \rightarrow \infty)$ from the NEMD simulations are in fact consistent with each other, as expected from linear response theory. Furthermore, we show that by multiplying the correlation time with a reasonable effective phonon group velocity, the EMD and NEMD data overlap each other very well. Our results thus firmly establish the equivalence between the two methods in different spatial dimensions, when the proper limits of long times and large system sizes are carefully considered.

\section{Models and Methods}

\begin{figure*}[htb]
\begin{center}
\includegraphics[width=2\columnwidth]{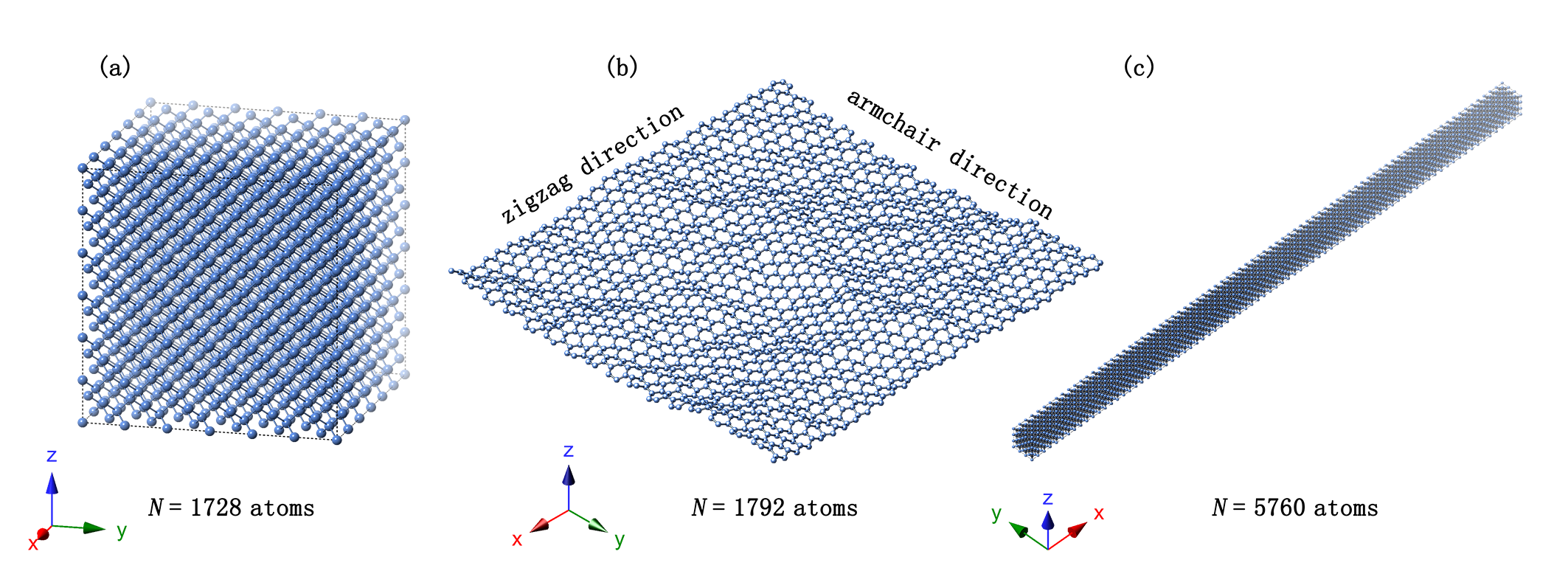}
\caption{Schematic illustration of the model systems studied in this work: (a) 3D bulk silicon; (b) 2D silicene; (c) Q1D SiNW. The cell size shown here for bulk silicon is the same as that used in the EMD simulations, but for clarity, the cell sizes for silicene and SiNW shown here are smaller than those used in the EMD simulations. In the NEMD simulations, the cell sizes in the transport direction ($x$ direction) can be much larger. See text and Table \ref{table:data} for details.}
\label{figure:models}
\end{center}
\end{figure*}

In this work, we use both the EMD and the NEMD methods for thermal conductivity calculations as implemented in the GPUMD package \cite{fan2017cpc, fan2017gpumd}. 

\subsection{Models}

We study three Si-based materials: 3D bulk silicon crystal, 2D silicene, and Q1D SiNW, which are schematically shown in Fig. \ref{figure:models}. For simplicity, we only consider isotopically pure systems although this is not a restriction of the methods used. We use classical MD simulations with empirical many-body potentials. For 3D bulk silicon, we chose to  use the Tersoff potential \cite{tersoff1989prb} with the original parameterization because a comprehensive comparison between the EMD and the NEMD methods has already been done by Howell \cite{howell2012jcp} using the SW potential \cite{stillinger1985prb}. For 2D silicene, we used the SW potential \cite{stillinger1985prb} re-parameterized by Zhang \textit{et al.} \cite{zhang2014prb}. To be consistent with Zhang \textit{et al.} \cite{zhang2014prb}, the thickness of single-layer silicene was chosen as 4.20 {\AA} when calculating the sample volume in the EMD method and the cross-sectional area in the NEMD method. Last, for Q1D SiNW, we used the SW \cite{stillinger1985prb} potential with the original parameterization, following Yang \textit{et al.} \cite{yang2010nt}. In all the MD simulations, we first equilibrated the system to room temperature and zero pressure conditions. Effects of temperature and external pressure were not considered here. 

Different boundary conditions were adopted for different model systems. In the EMD simulations, we used periodic boundary conditions in all the three directions for bulk silicon, the in-plane directions ($xy$ plane) of silicene, and the longitudinal direction ($x$ direction) of SiNW. Free boundary conditions were used for the out-of-plane direction in silicene and ripples formed automatically during the MD simulations (Fig. \ref{figure:models}(b)). For SiNW, we adopted fixed boundary conditions in the transverse directions ($y$ and $z$) in order to be consistent with the simulations by Yang \textit{et al.} \cite{yang2010nt}, although free boundary conditions can also be used. The fixed atoms are excluded in determining the volume and cross-sectional area. In the NEMD simulations, the two ends of system in the transport direction were fixed. 

The simulation cells were chosen as follows. For bulk silicon and SiNW, the coordinate axes were aligned along the [100] lattice directions. A simulation cell consisting of $N_x\times N_y \times N_z = 6\times6\times6$ conventional cubic cells with a total of $N=1728$ atoms was used for bulk silicon in the EMD simulations. In the NEMD simulations, we kept $N_y$ and $N_z$ unchanged and chose several values of $N_x$ such that the length $L_x$ varies from about 82 nm to 1 $\mu$m. For SiNW, we chose $N_y = N_z = 3$ and fixed the surface layer of atoms (same as in Ref. \cite{yang2010nt}) in both the EMD and the NEMD simulations. The length $L_x$ was chosen to be about 50 nm in the EMD simulations and was varied from 0.5 $\mu$m to 3 $\mu$m in the NEMD simulations. For silicene, the $x$ and $y$ axes pointed to the zigzag and armchair directions, respectively, and a roughly square-shaped simulation cell with $N=8640$ atoms was used in the EMD simulations. In the NEMD simulations, the width was kept to be about $L_y=10$ nm and the length $L_x$ was varied from about 40 nm to 320 nm. We checked that the cell sizes used in the EMD simulations were large enough to eliminate finite-size effects.

\subsection{The EMD method}

The EMD method for thermal conductivity calculations is based on the Green-Kubo formula \cite{green1954jcp,kubo1957jpsj}, which expresses the (running) thermal conductivity tensor $\kappa_{\mu\nu}(t)$ as an integral of the heat current autocorrelation function (HCACF) $\langle J_{\mu} (0)J_{\nu} (t) \rangle$ with respect to the correlation time $t$:
\begin{equation}
\label{equation:gk}
\kappa_{\mu\nu}(t) = \frac{1}{k_BT^2V}\int_0^{t} \langle J_{\mu}(0) J_{\nu}(t') \rangle dt'.
\end{equation}
Here, $k_B$ is the Boltzmann constant, $T$ is the absolute temperature of the system, $V$ is the volume, and $J_{\mu}$ is the heat current in the $\mu$ direction. Generally, one can obtain the whole conductivity tensor, but we are only interested in the diagonal elements here.

For many-body potentials such as the Tersoff and the SW potentials used in this work, the heat current $\vect{J}$ can be expressed as \cite{fan2015prb}
\begin{equation}
\vect{J} = \sum_i \sum_{j \neq i} \vect{r}_{ij}
\frac{\partial U_j}{\partial \vect{r}_{ji}} \cdot \vect{v}_i,
\end{equation}
where $\vect{r}_{ij} \equiv \vect{r}_{j} - \vect{r}_{i}$ and $\vect{r}_i$, $\vect{v}_i$, and  $U_i$ are respectively the position, velocity, and potential energy of atom $i$. Following Ref. \cite{fan2017prb}, we consider the in-out decomposition of the heat current for 2D systems, $\vect{J}=\vect{J}^{\rm in}+\vect{J}^{\rm out}$, where $\vect{J}^{\rm in}$ only includes the terms with $v_x$ and $v_y$ and $\vect{J}^{\rm out}$ only includes the terms with $v_z$. With this heat current decomposition, the running thermal conductivity along the $x$ direction can be naturally decomposed into three terms:
\begin{equation}
\label{equation:kappa_total}
\kappa_{x}(t)  =
\kappa^{\rm in}_{x}(t)+\kappa^{\rm out}_{x}(t)+\kappa^{\rm cross}_{x}(t),
\end{equation}
where
\begin{eqnarray}
\kappa^{\rm in}_{x}(t)&=&
\frac{1}{k_BT^2V} \int_0^{t} dt'
\langle J^{\rm in}_x(t') J^{\rm in}_x(0) \rangle ;  \\
\kappa^{\rm out}_{x}(t)&=&
\frac{1}{k_BT^2V} \int_0^{t} dt'
\langle J^{\rm out}_x(t') J^{\rm out}_x(0) \rangle ;
\\
\kappa^{\rm cross}_{x}(t)&=&
\frac{2}{k_BT^2V} \int_0^{t} dt'
\langle J^{\rm in}_x(t') J^{\rm out}_x(0) \rangle .
\end{eqnarray}

In the EMD simulations, we first equilibrated the system in the NPT ensemble with a temperature of $T=300$ K and a pressure of $p=0$ GPa for 2 ns.  After equilibration, we evolved the system for another 20 ns in the NVE ensemble and recorded the heat current data for later post-processing. We performed 50 independent simulations for each material to ensure sufficient statistics.

\subsection{The NEMD method}

\begin{figure}[htb]
\begin{center}
\includegraphics[width=\columnwidth]{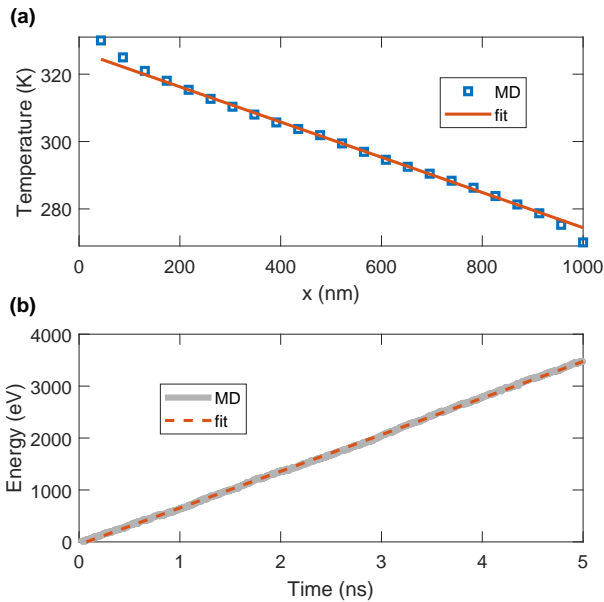}
\caption{(a) Steady-state temperature profile in the longest (1 $\mu$m) bulk silicon system. A linear fit to the block temperatures excluding a few blocks around the heat source and sink regions gives the absolute value of the temperature gradient $|\nabla T|$. (b) The energy of the thermostat (averaged over the source and the sink) as a function of the time in steady state. The heat transfer rate $dE/dt$ is calculated as the slope of the linear fit (dashed lines).}
\label{nemd}
\end{center}
\end{figure}

The NEMD method can be used to calculate the thermal conductivity $\kappa(L_x)$ of a system of finite length $L_x$ according to Fourier's law,
\begin{equation}
\label{Fourier}
\kappa(L_x) =
\frac{Q}{|\nabla T|},
\end{equation}
in the linear response regime where the temperature gradient $|\nabla T|$ across the system is sufficiently small. We generate the nonequilibrium steady-state heat flux $Q$ by coupling a source region of the system to a thermostat (realized by using the Nos\'{e}-Hoover chain method \cite{nose1984jcp,hoover1985pra,martyna1992jcp}) with a higher temperature of 330 K and a sink region to a thermostat with a lower temperature of 270 K. When steady state is achieved, the heat flux $Q$ can be calculated from the energy transfer rate $dE/dt$ between the source/sink and the thermostats:
\begin{equation}
\label{heat current}
Q = \frac{dE/dt}{S},
\end{equation}
where $S$ is the cross-sectional area perpendicular to the transport direction. Both the temperature gradient and the energy transfer rate were determined by linear fitting, as illustrated in Fig. \ref{nemd} for one independent simulation in the case of bulk silicon with a system length of 1 $\mu$m. Note that we reported the system length in the NEMD simulations as the source-sink distance, not excluding the regions with nonlinear temperature dependence around the source and sink, which was suggested to be a reasonable definition according to Howell \cite{howell2011jctn}.

In the NEMD simulations, we first equilibrated the system in the NPT ensemble ($T=300$ K and $p=0$ GPa) for $2$ ns and then generated the nonequilibrium heat current for $10$ ns. Steady state can be well achieved within $5$ ns, and we thus used the data during the later $5$ ns to determine the temperature gradient and the nonequilibrium heat current.  We performed five independent simulations for each system with a given length. In all the EMD and NEMD simulations, we used the velocity-Verlet integration scheme \cite{swope1982jcp} with a time step of $1$ fs, which has been tested to small enough.

\section{Results and discussion}

\begin{table*}[tbp]
\centering
\caption{Relevant data from the EMD and NEMD simulations: simulation cell length $L_x$ (in units of nm), number of atoms $N$ (including the fixed atoms),  the average thermal conductivity  $\kappa_{\rm ave}$ (in units of W m$^{-1}$ K$^{-1}$) from a number of independent simulations (50 in the EMD simulations and 5 in the NEMD simulations), and the standard error $\kappa_{\rm err}$. } 
\label{table:data}
\begin{tabular}{@{}r|rrrr|rrrr|rrrr|rrrr|rrrr@{}}
    \hline
    \hline
Method                              &
\multicolumn{4}{c}{Bulk silicon}    & 
\multicolumn{4}{c}{Bulk silicon}    &
\multicolumn{4}{c}{Silicene (SW1)}  &                         
\multicolumn{4}{c}{Silicene (SW2)}  & 
\multicolumn{4}{c}{SiNW} \\ 
\hline
                                                                 & 
$L_x $    & $N $    & $\kappa_{\rm ave}$  & $\kappa_{\rm err}$   &
$L_x $    & $N $    & $\kappa_{\rm ave}$  & $\kappa_{\rm err}$   &
$L_x $    & $N $    & $\kappa_{\rm ave}$  & $\kappa_{\rm err}$   & 
$L_x $    & $N $    & $\kappa_{\rm ave}$  & $\kappa_{\rm err}$   & 
$L_x $    & $N $    & $\kappa_{\rm ave}$  & $\kappa_{\rm err}$   \\
\hline
  
NEMD &  82  & 44064  &  61   &  1 &  327  & 176256  &  139   &  1  & 38   & 6528   & 8.4 & 0.4 & 38   & 6528   & 11.9 & 0.4  & 500  &  66312 & 40 & 1\\
 & 109  & 58752  &  75   &  1  &  490  & 264384  &  163   &  4 & 75   & 13056  & 8.8 & 0.2 & 75   & 13056  & 13.0 & 0.1  & 1000 & 132552 & 52 & 1\\
 & 136  & 73440 & 86   &  2  &  571  & 308448  & 177   &  1 & 150  & 26112  & 9.0 & 0.2 & 150  & 26112  & 13.2 & 0.3  & 1500 & 198792 & 58 & 2\\
 & 163  & 88128 & 95   &  1  &  653  & 352512  &  180   &  3 & 224  & 39168  & 9.0 & 0.2 & 225  & 39168  & 13.2 & 0.2  & 2000 & 265032 & 63 & 3\\ 
 & 245  & 132192 & 121   &  4  &  1000  & 529920  & 206   &  2 & 298  & 52224  & 9.1 & 0.2 & 300  & 52224  & 13.4 & 0.2  & 3000 & 397512 & 64 & 1\\
 \hline
EMD &  &&&&  & 1728   & 250   &  10  &   & 8640   & 9.3 & 0.1 &   & 8640   & 13.4 & 0.1  &   &   6624 & 65 & 2\\
\hline
\hline
\end{tabular}
\end{table*}

\subsection{3D bulk silicon}

\begin{figure}[hbt]
\begin{center}
\includegraphics[width=\columnwidth]{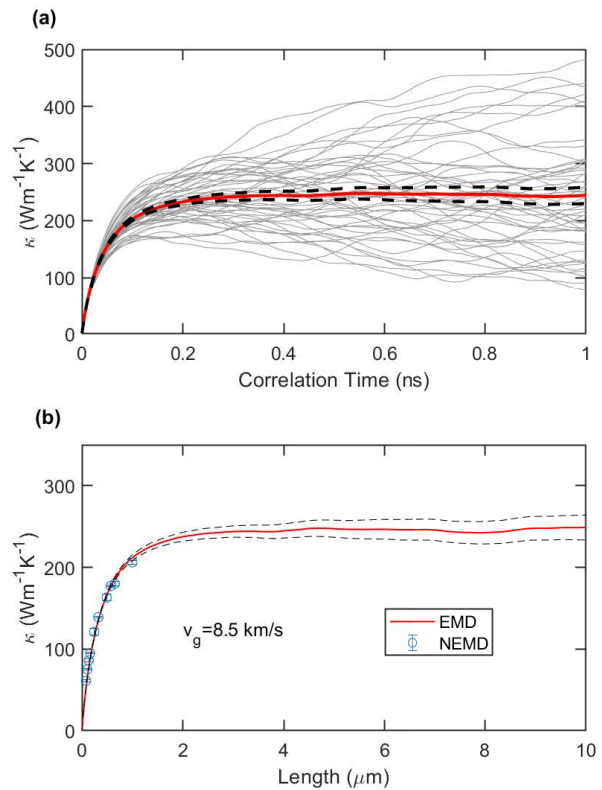}
\caption{(a) Running thermal conductivity for bulk silicon at 300 K and zero pressure as a function of correlation time. The thin lines represent the results from 50 independent simulations and the thick solid and dashed lines their average and error bounds.  (b) Thermal conductivity as a function of system length from EMD and NEMD simulations. An effective phonon group velocity of $v_g=8.5$ km/s was used to obtain the effective system length from the correlation time in the Green-Kubo formula. }
\label{figure:silicon}
\end{center}
\end{figure}

We start by discussing the results for bulk silicon. Figure \ref{figure:silicon}(a) shows the running thermal conductivities from 50 independent simulations as thin lines, each with a different set of initial velocities. The running thermal conductivity can vary from simulation to simulation and the variation increases with increasing correlation time, which means that the variation in the HCACF does not decay with increasing correlation time. This is a general property of time-correlation functions and transport coefficients in MD simulations \cite{haile1992book}. The average $\kappa_{\rm ave}(t)$ of the independent runs is shown as a thick solid line in Fig. \ref{figure:silicon}(a). To quantify the error bounds, we calculated the standard error $\kappa_{\rm err}(t)$ (standard deviation divided by the square root of the number of simulations) and plot $\kappa_{\rm ave}(t) \pm \kappa_{\rm err}(t)$ as dashed lines. It can be seen that $\kappa_{\rm ave}(t)$ converges well in the time interval $[0.5 ~{\rm ns}, 1 ~{\rm ns}]$. By averaging $\kappa_{\rm ave}(t)$ and $\kappa_{\rm err}(t)$ within this range, we finally get an average value of the thermal conductivity and its error estimate: $\kappa_{\rm ave} \pm \kappa_{\rm err} = 250 \pm 10$ W m$^{-1}$ K$^{-1}$. These and other relevant data are summarized in Table \ref{table:data}. 

Figure \ref{figure:silicon}(b) shows the NEMD results as markers with error bars, representing respectively the average and the standard error from five independent simulations for each system length. The same data are listed in Table \ref{table:data}. It can be seen that $\kappa$ calculated from the NEMD simulations increases with increasing length, which is a sign of ballistic-to-diffusive transition. Similar information is incorporated in the running thermal conductivity from the EMD simulations. Actually, we can make closer comparisons between the EMD and the NEMD results. One can define an effective system length $L_x$ in the EMD method by multiplying the upper limit of the correlation time $t$ in the Green-Kubo formula Eq. (\ref{equation:gk}) by an effective phonon group velocity $v_{\rm g}$:
\begin{equation}
\label{equation:t2l}
L_{x} \approx v_{\rm g} t.
\end{equation}
The running thermal conductivity $\kappa(t)$ in the EMD method can also be regarded as a function of the system length $\kappa(L_x)$, which can be directly compared with the NEMD results. The concept of effective phonon group velocity has been extensively used in the study of heat transport in low-dimensional lattice models \cite{lepri2003pr} and has also been recently used for graphene \cite{fan2017prb}. By treating $v_{\rm g}$ as a free parameter, we can obtain a good match between the EMD and the NEMD data, as shown in Fig. \ref{figure:silicon}(b). This effective group velocity is by no means to be taken as a quantitatively accurate value for the average phonon group velocity, because Eq. (\ref{equation:t2l}) is not an exact expression. We consider a set of candidate solutions of the group velocity with an interval of 0.1 km s$^{-1}$ and choose the group velocity value which gives the smallest difference between the NEMD and EMD data at appropriate points. Nonetheless, the fitted value, $v_{\rm g} = 8.5$ km s$^{-1}$, is comparable to the longitudinal ($8.69$ km s$^{-1}$) and transverse ($5.28$ km s$^{-1}$) acoustic phonon group velocities calculated using density functional theory \cite{marchbanks2015jap}. The important result here is that the length-convergence trends of thermal conductivity from both EMD and NEMD simulations are consistent with each other. To fully demonstrate the consistency between the two methods, we would need to consider longer systems (up to several microns) in the NEMD simulations, which is computationally prohibitive for bulk silicon.

\begin{figure}[hbt]
\begin{center}
\includegraphics[width=\columnwidth]{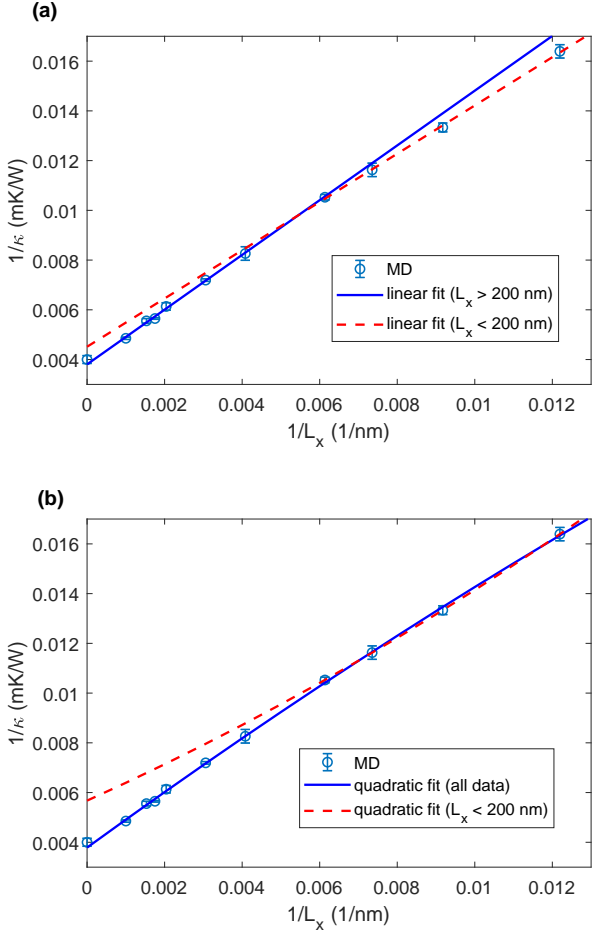}
\caption{Inverse thermal conductivity $1/\kappa$ as a function of the inverse simulation cell length $1/L_x$ in the NEMD simulations. In both (a) and (b), the markers are MD data, with the one at $1/L_x=0$ representing the value from EMD and others from NEMD. In (a), the solid and dashed lines represent linear fits (using Eq. (\ref{equation:fit1})) to the NEMD data with $L_x>200$ nm and $L_x<200$ nm, respectively. In (b), the solid and dashed lines represent quadratic fits (using Eq. (\ref{equation:fit2})) to all the NEMD data and the NEMD data $L_x<200$ nm, respectively. See text for details.}
\label{figure:extrapolation}
\end{center}
\end{figure}

One way to explore the consistency between the two methods based on a finite amount of NEMD data is to extrapolate the conductivity values of finite systems to the limit of infinite length using certain empirical expressions. The simplest extrapolation formula is the one proposed by Schelling \textit{et al.} \cite{schelling2002prb}:
\begin{equation}
\label{equation:fit1}
\frac{1}{\kappa(L_x)} = \frac{1}{\kappa_0} \left(1 + \frac{\lambda}{L_x} \right),
\end{equation}
where $\kappa_0=\kappa(L_x \to \infty)$ is the extrapolated thermal conductivity in the infinite-length limit and $\lambda$ is an effective phonon mean free path that is conceptually similar to the effective phonon group velocity defined by Eq.~(\ref{equation:t2l}). This is a first-order expression which is only good when the system lengths are comparable or larger than the effective phonon mean free path \cite{sellan2010prb}. With a wide range of system lengths, the thermal conductivity data usually exhibit a nonlinear relation between $1/\kappa(L_x)$ and $1/L_x$. Figure \ref{figure:extrapolation} (a) shows that a linear fit to the NEMD data with $L_x>200$ nm results in an extrapolated thermal conductivity of $\kappa_0 = 260\pm 10$ W m$^{-1}$ K$^{-1}$, which is consistent with the EMD value. In contrast, a linear fit to the NEMD data with $L_x<200$ nm results in a value of $\kappa_0 = 220\pm 10$ W m$^{-1}$ K$^{-1}$, which is appreciably smaller than the EMD value. The effective phonon mean free path is determined to be $\lambda\approx 300$ nm, which explains why an inaccurate $\kappa_0$ is obtained using the NEMD data with $L_x<200$ nm. Figure \ref{figure:extrapolation} (b) shows that the nonlinear behavior can otherwise be well described by a second-order expression \cite{sellan2010prb,zaoui2016prb}
\begin{equation}
\label{equation:fit2}
\frac{1}{\kappa(L_x)} = \frac{1}{\kappa_0} \left(1 + \frac{\lambda}{L_x} + \frac{\beta}{L_x^2} \right),
\end{equation}
where $\beta$ is a parameter of the dimension of length squared. Alternatively, the nonlinearity may also be captured by expressions with fractional powers of $1/L_x$ \cite{allen2014prb}. However, when using the NEMD data with $L_x<200$ nm, the quadratic fit also fails to yield the correct extrapolated $\kappa_0$ (cf. the dashed line in Fig. \ref{figure:extrapolation} (b)). Therefore, no matter what expression is used in the fit, using NEMD data with relatively short simulation cell lengths may result in significant errors and is a possible reason for some reported inconsistencies between the EMD and NEMD methods. Recently, Liang \textit{et al.} \cite{liang2015jap} found that the extrapolated $\kappa_0$ obtained by using the linear fit to their NEMD data with $L_x \leq 150$ nm is $166\pm 11$ W m$^{-1}$ K$^{-1}$ for bulk GaN (described by a SW potential) at 300 K, which is several times smaller than their EMD value, $1190\pm 85$ W m$^{-1}$ K$^{-1}$. They also attributed the inconsistency between the NEMD and EMD predictions to the inadequacy of the linear extrapolation.

\subsection{2D silicene}

\begin{figure}[hbt]
\begin{center}
\includegraphics[width=\columnwidth]{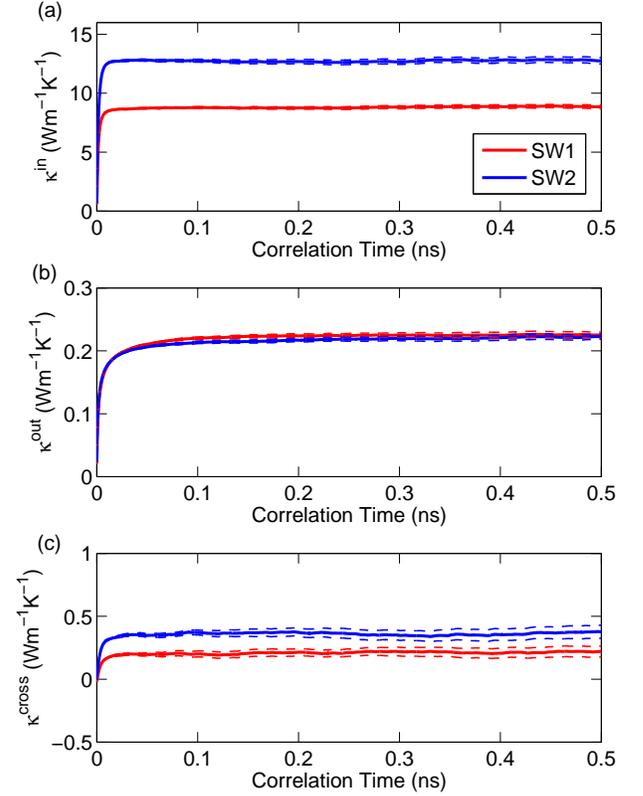}
\caption{Running thermal conductivity for silicene at 300 K and zero pressure as a function of correlation time for the (a) in-plane component, (b) the out-of-plane component, and (c) the cross-term. The red and blue lines correspond to the results obtained by using the SW1 and the SW2 parameter sets, respectively. The solid and dashed lines respectively represent the averages and the standard errors from 50 independent runs.}
\label{figure:emd01}
\end{center}
\end{figure}

We next consider 2D silicene. Figure \ref{figure:emd01} shows the running thermal conductivity components, $\kappa^{\rm in}$, $\kappa^{\rm out}$, and $\kappa^{\rm cross}$, using the two SW parameter sets given by Ref. \cite{zhang2014prb}. We checked that there is no noticeable difference between $\kappa_x$ and $\kappa_y$, which means that the system is isotropic in terms of heat transport. In view of this, we report the average $\kappa=(\kappa_x+\kappa_y)/2$ in Fig. \ref{figure:emd01}. The dashed lines in Fig. \ref{figure:emd01} indicate standard errors calculated from 50 independent simulations, similar to the case of bulk silicon. 

All the running thermal conductivity components well converge within a fraction of a nanosecond, faster than the case of bulk silicon. The converged total thermal conductivity value is also significantly smaller than that in bulk silicon.  The parameter set SW1 gives noticeably smaller $\kappa^{\rm in}$, while both parameter sets give comparable $\kappa^{\rm out}$. For each parameter set, $\kappa^{\rm in}$ converges to a much higher value than $\kappa^{\rm out}$ does, which is opposite to the case of graphene \cite{fan2017prb}. It is also interesting to note that $\kappa^{\rm cross}$ does not converge to zero, which can be understood by the fact that there is intrinsic corrugation in silicene, similar to the case of polycrystalline graphene \cite{fan2017nl}. Based on visual inspection, we chose the time interval $[0.3~\text{ns} - 0.5~\text{ns}]$ to evaluate the converged thermal conductivity, which was determined to be $9.3 \pm 0.1$ W m$^{-1}$ K$^{-1}$ and $13.4 \pm 0.1$ W m$^{-1}$ K$^{-1}$, respectively, for the SW1 and SW2 parameter sets.

\begin{figure}[hbt]
\begin{center}
\includegraphics[width=\columnwidth]{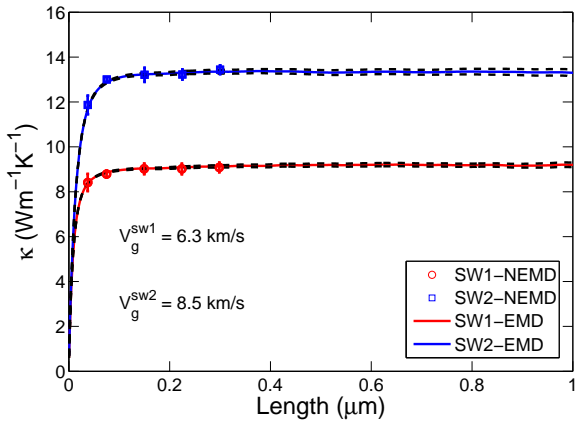}
\caption{(a) Direct comparison between NEMD (markers) and EMD (lines) data. For the EMD data, the system length is calculated from the correlation time according to Eq. (\ref{equation:t2l}). See text for details. }
\label{silicene-nemd}
\end{center}
\end{figure}

The NEMD results for silicene are shown in Fig. \ref{silicene-nemd}. We can obtain a good match between the EMD and the NEMD data for both parameter sets, with the effective group velocities being fitted to be  $6.3$ km s$^{-1}$ and $8.5$ km s$^{-1}$, respectively. The ratio between the effective group velocities from the two parameter sets is close to that between the thermal conductivities. The fact that the SW1 parameter set gives a smaller effective phonon group velocity can also be confirmed by examining the phonon dispersions given in Ref. \cite{zhang2014prb}. In Ref. \cite{zhang2014prb}, it was found that the EMD method gives significantly smaller $\kappa$ than the NEMD method, which put the consistency between the two methods into question. However, our results unequivocally show that the two methods give consistent results for both parameter sets. The reason for the inconsistency in the previous work is that the heat current formula as implemented in the LAMMPS code \cite{plimpton1995jcp,lammps} used in Ref. \cite{zhang2014prb} is not applicable to many-body potentials such as the SW potential, as pointed out in Ref. \cite{fan2015prb} and further demonstrated in Ref. \cite{gill2015prb}. In contrast, the heat current formula as implemented in the GPUMD code \cite{fan2017cpc,fan2017gpumd} used in the current work has been fully validated \cite{gill2015prb,fan2017prb}.

\subsection{Q1D silicon nanowire}

\begin{figure}[hbt]
\begin{center}
\includegraphics[width=\columnwidth]{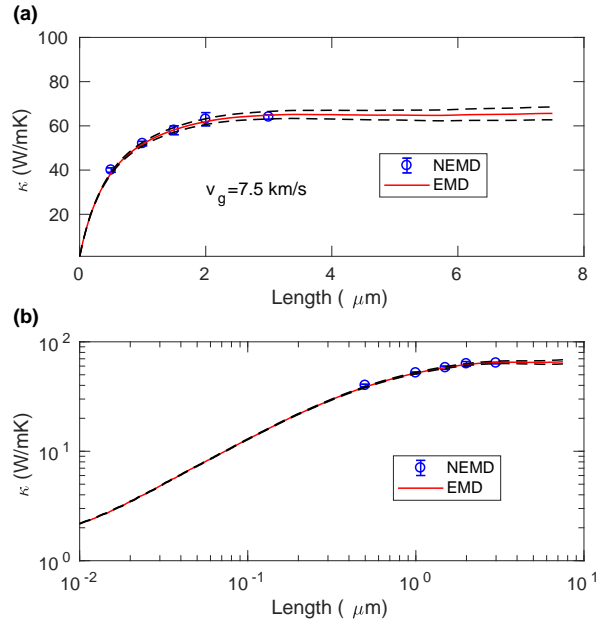}
\caption{Thermal conductivity as a function of system length from EMD and NEMD simulations with the axes in normal (a) and log-log (b) scales. An effective phonon group velocity of $v_g=7.5$ km/s was used to obtain the effective system length from the correlation time in the Green-Kubo formula. }
\label{figure:1d}
\end{center}
\end{figure}

Last, we consider Q1D SiNW. Figure \ref{figure:1d}(a) shows the thermal conductivity values from EMD and NEMD simulations as a function of system length, where an effective phonon group velocity of $v_{\rm g}=7.5$ km s$^{-1}$ was used to convert the correlation time to an effective system length in the EMD method. Because the cross-sectional area used here is much smaller than that used in the case of bulk silicon, we have reached a longer system of length 3 $\mu$m in the NEMD simulations. At this length, we obtain a thermal conductivity of $64 \pm 1$ W m$^{-1}$ K$^{-1}$, which agrees with the converged value from the EMD simulations, $65 \pm 2$ W m$^{-1}$ K$^{-1}$. This suggests that the two methods gives consistent results and ultra-thin SiNW with fixed boundaries in the transverse directions has much smaller converged thermal conductivity than that of the bulk silicon. Yang \textit{et al.} \cite{yang2010nt} reported a power-law divergent thermal conductivity with respect to the system length based on their NEMD data. Our results do not support this viewpoint. In Fig. \ref{figure:1d}(b), we plot the same data from Fig. \ref{figure:1d}(a) but with a log-log scale. There might be a region where one can make a power-law fit, but the thermal conductivity eventually converges to a finite value.

\section{Summary and Conclusions}

In summary, we have compared the EMD and NEMD methods for computing thermal conductivity in three Si-based systems with different spatial dimensions: 3D bulk silicon, 2D silicene, and Q1D SiNW. Particularly, by converting the correlation time in the EMD method to an effective system length according to Eq. (\ref{equation:t2l}) with an appropriate value of the effective phonon group velocity, we can compare the EMD results directly with the NEMD results. For all the systems, we found excellent agreement between the two methods. While it is computationally prohibitive to directly obtain length-convergent thermal conductivity in the case of bulk silicon, we achieved this for silicence and SiNW, where the length-convergent thermal conductivities from the NEMD method were found to be consistent with the time-converged thermal conductivities from the EMD method. Our results thus firmly establish the expected equivalence between the two methods when long enough times and large enough systems are used in the simulations. We also note that some of the discrepancies reported in the literature are due to an incorrect implementation of the heat current for many-body potentials in LAMMPS. Inappropriate use of the linear extrapolation as expressed by Eq. (\ref{equation:fit1}) is another possible cause of inconsistency between the two methods.

\begin{acknowledgments}
This work was supported in part by the National Natural Science Foundation of China under Grant No. 11404033 and  
in part by the Academy of Finland Centre of Excellence program (project 312298). We acknowledge the computational resources provided by Aalto Science-IT project and Finland's IT Center for Science (CSC).
\end{acknowledgments}

\bibliography{refs}

\end{document}